\documentclass[%
 reprint,
superscriptaddress,
amsmath,amssymb,
prb,
]{revtex4-1}

\usepackage{graphicx}
\usepackage{dcolumn}
\usepackage{bm}
\usepackage{amsmath}

\newcommand*{\citen}[1]{%
  \begingroup
    \romannumeral-`\x 
    \setcitestyle{numbers}%
    \cite{#1}%
  \endgroup   
}

\DeclareMathAlphabet\mathbfcal{OMS}{cmsy}{b}{n}  

\begin{document}

\preprint{APS/123-QED}

\title{Magnetoelectric behavior of breathing kagom\'{e} monolayers of $\mathrm{Nb}_3\mathrm{(Cl,Br,I)}_8$ from first-principles calculations} 
%
%

\author{John Mangeri}
 \email{johnma@dtu.dk}
 \author{Varun R. Pavizhakumari}
 \author{Thomas Olsen}
 \email{tolsen@fysik.dtu.dk}
\affiliation{%
Computational Atomic-scale Materials Design (CAMD), Department of Physics, Technical University of Denmark, \\ 
 Anker Engelunds Vej 101 2800 Kongens Lyngby, Denmark\\
}%
\date{\today}

\begin{abstract}

We apply density functional theory to explore the magnetoelectric (ME) properties of two-dimensional $\mathrm{Nb}_3\mathrm{(Cl,Br,I)}_8$. These compounds have recently been proposed to exhibit coupled ferroelectric and ferromagnetic order leading to a switchable anomalous valley Hall effect (AVHE).
Using both spin-spiral and self-consistent spin-orbit coupled calculations, we predict an in-plane $120^\circ$ cycloid of trimerized  spins as the ground state for $\mathrm{Nb}_3\mathrm{Cl}_8$.
For $\mathrm{Nb}_3\mathrm{Br}_8$ and $\mathrm{Nb}_3\mathrm{I}_8$ we find long period incommensurate helical order.
We calculate a number of magnetic properties such as the exchange constants, orbital magnetization, and Weiss temperatures.
It is then shown that, despite having both broken inversion and time-reversal symmetry, the proposed AVHE and linear ME response are forbidden by the presence of helical order in the ground state.
In addition, the computed switching trajectory demonstrates that it is unlikely that the polar state of the monolayers can be switched with a homogeneous electric field due to an unusual equation of state of the out-of-plane dipole moment.
Nevertheless, we highlight that in the presence of a strong electric field, the trimerized spins in $\mathrm{Nb}_3\mathrm{Cl}_8$ will exhibit a magnetic phase transition from the $120^\circ$ cycloid to out-of-plane ferromagnetic order, which restores the symmetry required for both AVHE and linear ME effects.
\end{abstract}

\maketitle


\section{\label{sec:intro} Introduction}

\vspace*{-10pt}

The niobium halide compounds $\mathrm{Nb}_3\mathrm{X}_8$ with X = (Cl, Br, I) have recently received attention due the presence of topologically flat bands, possible quantum spin liquid ground state and idealized Mott-Hubbbard physics \cite{Haraguchi2017, Jiang2017, Pasco2019, Conte2020, Peng2020, Sun2022, Regmi2022, Regmi2023, Wang2023, Feng2023, Zhang2023, Grytsiuk2024, Haraguchi2024, Liu2024}.
The structure is composed of van der Waals bonded monolayers, each of which is constructed from triangular units of $[\mathrm{Nb}_3]^{8+}$ condensed into a breathing kagom\'{e} lattice with alternating contracting and expanding Nb-Nb bonds. The two-dimensional (2D)  kagom\'{e} lattice is passivated by 8 halogen atoms (Cl, Br or I) situated above and below the Nb plane.  
Each $\mathrm{Nb}_3$ cluster contributes a single unpaired electron \emph{trimerized} as a $2a_1$ molecular orbital\cite{Haraguchi2017, Grytsiuk2024}, which constitutes an isolated spin-1/2 site.
Such multimer configurations are common across different  materials groups and has been reported in the kagom\'{e}rized cluster magnets $\mathrm{LiZn}_2\mathrm{Mo}_3\mathrm{O}_8$, $\mathrm{Na}_4\mathrm{Ir}_3\mathrm{O}_8$, and $\mathrm{Na}_3\mathrm{Sc}_2\mathrm{Mo}_5\mathrm{O}_{16}$ which have shown signatures consistent with the current picture of quantum spin liquids\cite{Okamoto2007, Sheckelton2012, Chen2022}.
The bulk layered compound $\mathrm{Nb}_3\mathrm{Cl}_8$ exhibits a structural transition at 
$T\simeq 90$ K where the symmetry is reduced from $P\bar{3}m1$ to $C2/m$ and the transition is accompanied by loss of magnetic order
\cite{Haraguchi2017, Pasco2019, Sheckelton2017, Sun2022, Gao2023}.
This has been assigned to the emergence of charge disproportionation between individual layers that renders the ground state in a spin singlet
\cite{Haraguchi2017, Haraguchi2024}.
Thus, at low temperature the unpaired $S = 1/2$ electrons of each layer reconfigure to pair up in every other atomic plane thus quenching the magnetism in the bulk.
In addition, angle-resolved photoemission spectroscopy (ARPES) has shown that the bands are topologically flat \cite{Sun2022, Regmi2022, Regmi2023} and these materials are regarded as ideal platforms for studying Mott-Hubbard physics\cite{Gao2023, Hu2023, Grytsiuk2024} and emergent correlated phenomena\cite{Meng2024}.
Finally, there have been a number of proposals for using these materials in novel device concepts such as negative photoconductivity detectors\cite{Lee2024} or bipolar field effect transistors\cite{Lu2024} exploiting the fact that externally applied fields may be able to drive useful (and biocompatible\cite{Jiang2017}) functionalities.
%
%

%
The case of Nb$_3$Cl$_8$ has been exfoliated to the monolayer limit and demonstrated to be stable under ambient conditions\cite{Sun2022}. If the exfoliation is carried out above the structural transition it is not possible for an exfoliated monolayer to exhibit charge disproportionation upon cooling and a 2D magnetic ground state should be attainable at low temperatures. Moreover, the bulk structures exhibit space groups of either $C2/m$ or $P\bar{3}m1$, which are both centrosymmetric while the individual layers reside in the polar space group $P3m1$. Monolayers thus open for a range of possible effects due to the lack of inversion symmetry. For example the anomalous valley Hall effect (AVHE), which may then couple to magnetic order and ground state spontaneous polarization\cite{Peng2020, Feng2023}.
%
%
%
%

%
Most first principles investigations of these compounds have been carried out in the framework of density functional theory (DFT) where an out-of-plane FM ordering of the $\mathrm{Nb}_3$ trimer spin was assumed\cite{Jiang2017, Peng2020, Conte2020, Sun2022, Regmi2022, Regmi2023, Feng2023}.
While these studies reproduce the weakly dispersive bands and other properties compatible with experimental measurements on the bulk, a systematic unveiling of the magnetic ground states still seems to be lacking. 
However, since the electronic structure could be strongly dependent on the magnetic ground state, this may have a prominent influence on the proposed AVHE\cite{Peng2020, Feng2023} and other delicate features of magnetoelectric (ME) origin.

Recent dynamical mean field theory calculations by Grystiuk \emph{et al}\cite{Grytsiuk2024} suggest that the magnetic ground state of a monolayer could be comprised of trimerized in-plane spin spirals.
This is corroborated by a DFT study by a Hu \emph{et} \emph{al} \cite{Hu2024} in $\mathrm{W}_3\mathrm{Cl}_8$ monolayers demonstrating the 5d trimerized spins stay ferromagnetically aligned but a long range $120^\circ$ ordering forms between the $\mathrm{W}_3$ clusters. Finally the observed negative Weiss temperature in Nb$_3$Cl$_8$ provides a strong indication of the presence of antiferromagnetic exchange interactions\cite{Sheckelton2017, Pasco2019, Sun2022, Gao2023, Liu2024}.
In addition to magnetic properties, the polar ground state has been proposed to be switchable by an applied electric field, which would render isolated monolayers ferroelectric. First principles calculations of switching paths have been reported in these compounds\cite{Li2021, Feng2023, Bhardwaj2023, Hu2024, Zhao2024} utilizing the nudged elastic band (NEB) method and the barrier has been shown to be comparable to known 2D ferroelectrics\cite{Kruse2023}, which strongly imply that the Nb halides should be ferroelectric as well.
However, the energy along the switching path were provided as a function of a certain (arbitrary) reaction coordinate, which is not directly related to the polarization or an applied electric field. A detailed prediction of switching requires one to calculate the coercive field - for example from the electrical enthalpy. The crucial point is that electric fields couple to polarization rather than reaction coordinates and it is vital to express the energy as a function of polarization in order to demonstrate switching.
The theoretical description of possible switching mechanisms in these compounds thus remains rather incomplete.
In light of the considerations above, the present work provides a fresh look at niobium halide monolayers and investigate their magnetic properties as well as the possibility of switching the polarization by an electric field. 
Using the local spin density approximation (LDA) of DFT, we identify a $120^\circ$ spin-cycloid as the ground state of $\mathrm{Nb}_3\mathrm{Cl}_8$ while both $\mathrm{Nb}_3\mathrm{Br}_8$ and $\mathrm{Nb}_3\mathrm{I}_8$ host incommensurate long wavelength spin spirals.
We then calculate the exchange constants by fitting to a Heisenberg model and we extract the Weiss temperatures and compare with experimental bulk values.
%
%
The symmetry of the spirals are demonstrated to restore the reciprocity in $\mathrm{Nb}_3\mathrm{Cl}_8$ thus \emph{removing} the AVHE that was previously predicted based on a FM ground state\cite{Peng2020, Feng2023}. The presence (or lack of) AVHE thus comprises a highly useful signature of the magnetic ground state that may be probed directly by ARPES measurements.
We also probe the linear ME response and show that the cycloidal order both changes the symmetry and the intrinsic coupling of the ME tensor.

Finally, we raise the question of how the multiferroic order modulates or switches under an applied homogeneous electric field. 
We perform NEB calculations revealing the nonpolar transition state and a switching mechanism related to the breathing kagom\'{e} pattern as demonstrated in previous publications\cite{Feng2023, Hu2024, Zhao2024}.
However, despite having a transition barrier similar to other known 2D ferroelectrics, it is shown that additional zero points in the polarization profile along the switching path renders all the compounds non-switchable.
Nevertheless, even if the monolayers are not switchable in the usual sense of ferroelectrics, we show that the polarization may be strongly modified by an external electric field and that a magnetic phase transition to FM order accompanies such change. Thus, the AVHE and linear ME can be restored by application of an electric field thus demonstrating unconventional magnetoelectric behavior.

\vspace*{-10pt}

\section{\label{sec:ground} Ground states and magnetic properties}

\vspace*{-10pt}

\subsection{\label{subsec:structure} Structure}

\vspace*{-10pt}

\begin{figure}[t!]\centering
\includegraphics[height=7.1cm]{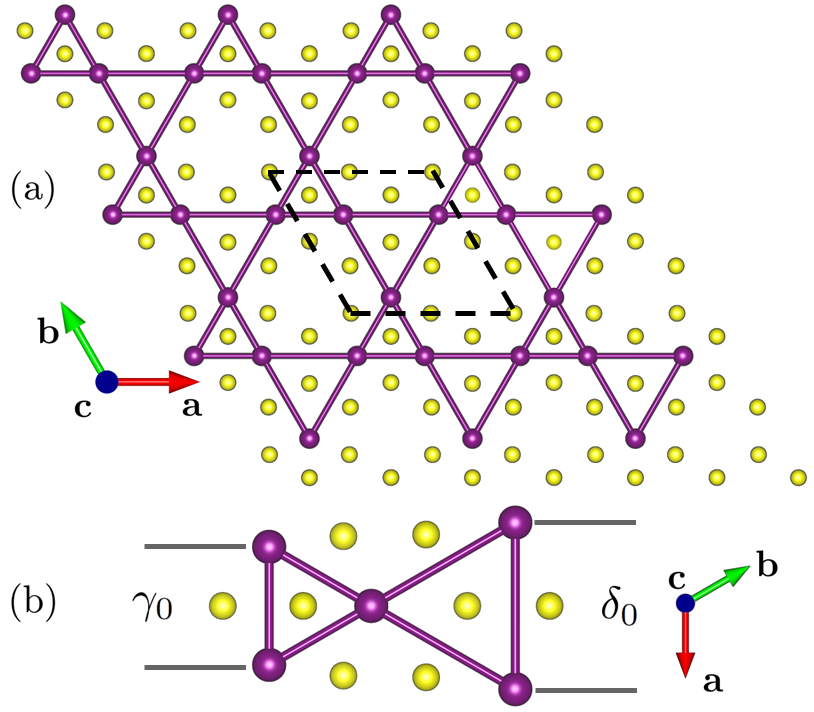} 
\caption{\label{fig1} (a) Ground state structure of $\mathrm{Nb}_3\mathrm{X}_8$, X = (Cl, Br, I) monolayers. Niobium is purple and halogens are colored in yellow. The dashed line shows the primitive cell of the formula unit. (b) Breathing kagom\'{e} lattice parameters $\gamma_0$ and $\delta_0$ corresponding to unequal bond lengths in adjacent corner-sharing equilateral $\mathrm{Nb}_3$ triangles.
}
\end{figure}

For structural relaxations, we perform collinear calculations of monolayers comprised of the formula unit $\mathrm{Nb}_3\mathrm{X}_8$ with $\mathrm{X} = (\mathrm{Cl}, \mathrm{Br}$, and $\mathrm{I})$ assuming an out-of-plane FM order of the $\mathrm{Nb}$ sites.
We work within the projector augmented wave (PAW) formalism\cite{Blochl1994} of DFT implemented in the \textsc{gpaw} electronic structure software package\cite{Mortensen2024} using the Atomic Simulation Environment\cite{Larsen2017}. 
Explicitly, 13 and 7 $e^-$ valence configurations of $4s^25s^14p^64d^4$, $3s^23p^5$, $4s^24p^5$ and $5s^25p^5$ are treated with PAW cutoff radii of $2.5, 1.5, 2.1,$ and $2.2$ \AA$\,$ for Nb, Cl, Br, and I respectively. 
We choose a plane-wave cutoff energy of 700 eV and relax the structures until the forces on all atoms are below $5$ meV/$\mathrm{\AA}$.
The reciprocal space is sampled with $\Gamma$-centered meshes of $12\times 12\times 1$ and $6\times 6\times 1$ $k$-points for primitive and supercells respectively and we applied the Perdew-Erhenzhof-Burke (PBE) exchange correlation functional \cite{Perdew1996}.
In order to isolate the structure from its periodic images we use a 
distance of 24 \AA $\,$ between periodic repetitions.
A dipole-layer correction is included \cite{Bengtsson1999} which ensures vanishing electric fields far from the monolayer despite the periodic boundary conditions and potential shift induced by the out-of-plane dipole moment.
Our calculations yield a set of relaxed in-plane lattice constants similar to what has been reported in the Computational Two-dimensional Materials Database (C2DB)\cite{Haastrup2018, Gjerding2021} and other publications\cite{Peng2020, Feng2023}.
The ground states obtained are all in the polar space group of $P3m1$ (No. 156) and exhibit out-of-plane polarization $P_z$.
As visualized\cite{Momma2011} in Fig.~\ref{fig1}~(a), the kagom\'{e} pattern is distorted (breathing) due to alternating corner-sharing small and large $\mathrm{Nb}_3$ triangles.
The Nb-Nb bonds can therefore be parsed into two parts with $\gamma_0$ (small) or $\delta_0$ (large) as exemplified in Fig.~\ref{fig1} (b) and with the general observation that $\delta_0 + \gamma_0 = a$ with $a$ the in-plane lattice constant.
The physical mechanism of the polarization can be associated with the halogen placements with respect to the relative orientation of the kagom\'{e} lattice.
Therefore, if the halogen appears above (below) the large triangle, we find a positive (negative) oriented dipole moment.
We summarize the calculated ground state structural data in Table \ref{table:tab1}.
\begin{table}[t!]
\begin{ruledtabular}
\begin{tabular}{c c c c c}
    & $\mathrm{Nb}_3\mathrm{Cl}_8$ & $\mathrm{Nb}_3\mathrm{Br}_8$  &  $\mathrm{Nb}_3\mathrm{I}_8$ \\
      a       & 6.802 & 7.155 &  7.684 & \\
   $\delta_0$ & 3.981 & 4.261 &  4.664 & \\  
   $\gamma_0$ & 2.821 & 2.894 &  3.020 & \\
   $P_z$ & -0.543 & -0.886 & -1.101 & \\
\end{tabular}
\end{ruledtabular}
\caption{\label{table:tab1}%
Lattice constants ($a$), large ($\delta_0$) and small ($\gamma_0$) breathing kagom\'{e} trimer bond lengths of the ground states of the niobium halides used in this work. Units are given in $\mathrm{\AA}$. The out-of-plane 2D polarization $P_z$ is calculated from the FM state and stated in units of $\mathrm{pC}\cdot\mathrm{m}^{-1}$.
}
\end{table}
\subsection{\label{subsec:mag} Magnetic order}
\begin{figure}[t!]\centering
\includegraphics[height=5.2cm]{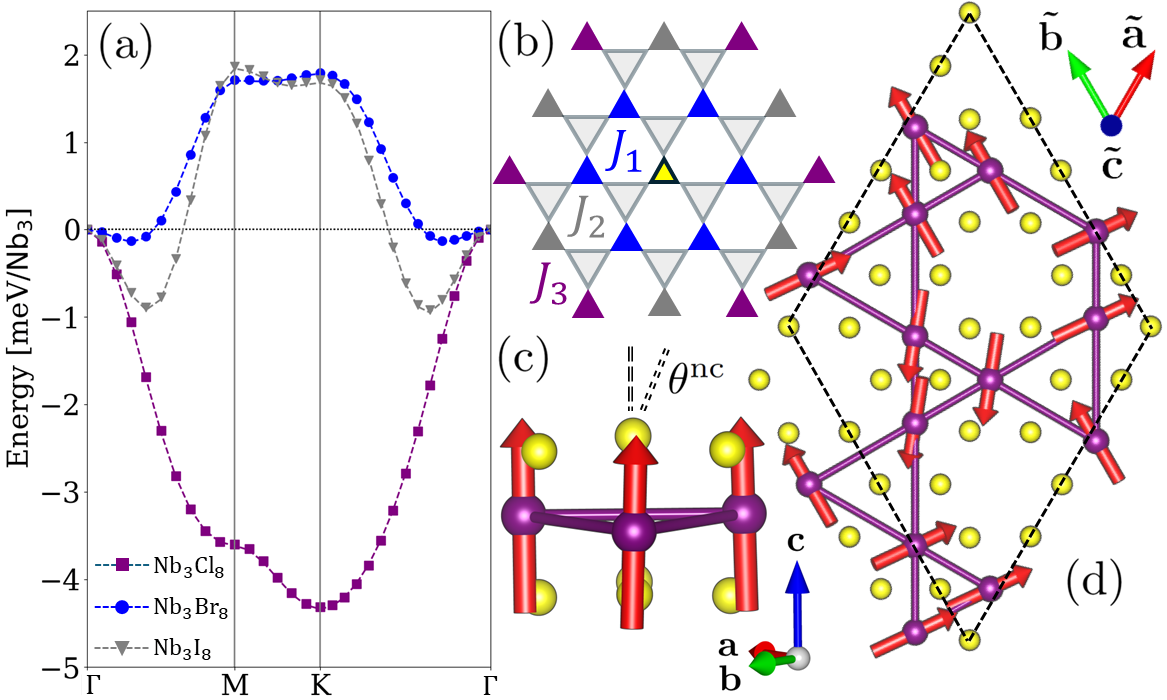} 
\caption{\label{fig2} (a) Spin-spiral energies as a function of ordering vector for $\mathrm{Nb}_3\mathrm{X}_8$ with X = Cl (purple squares), Br (blue circles) and I (gray triangles). The fits to a $J_1$-$J_2$-$J_3$ Heisenberg model shown in (b) are shown as dashed lines in (a). The FM configuration with weak non-collinear canting $(\theta^\mathrm{nc})$ found from DFT calculations including SOC is shown in in (c).
In (d) we demonstrate the ground state (coplanar) helical spin spiral of $\mathrm{Nb}_3\mathrm{Cl}_8$ with $\mathbf{Q} = (1/3, 1/3)$ obtained from the spin spiral calculations.}
%
\end{figure}

Next, we turn to the magnetic properties of the halides.
For this, the local density approximation (LDA) of the exchange correlation functional is used with our PBE relaxed structures.
To identify the magnetic ground states, we utilize the generalized Bloch theorem \cite{Knopfle2000,Heide2009,Zimmermann2019,Sødequist2023,Sodequist2024} to perform spin-spiral calculations of the three systems varying the ordering vector $\mathbf{q}$ along high symmetry lines of the primitive Brillouin zone.
The results are shown in Fig.~\ref{fig2}~(a) for $\mathrm{Nb}_3\mathrm{Cl}_8$ (purple squares), $\mathrm{Nb}_3\mathrm{Br}_8$ (blue circles) and $\mathrm{Nb}_3\mathrm{I}_8$ (gray triangles).
The FM state corresponds to $\mathbf{q}$ at the zone center, which is clearly not the ground state for any of the compounds.
In the case of $\mathrm{Cl}$, the minimum is located at the point K, which is represented as $\mathbf{Q} = (\frac{1}{3}, \frac{1}{3})$ (in reduced coordinates) indicating a $120^\circ$ cycloid configuration of the trimer spin.
For $\mathrm{Nb}_3\mathrm{(Br,I)}_8$, the minima are shifted slightly away from the $\Gamma$ point but are not found to be located at any of the high symmetry points indicating that a long wavelength incommensurate cycloid is favored.
It should be noted that the spin densities (and thus the three Nb spins) are relaxed during the spin spiral calculations, which only impose boundary conditions on the total spin density. We do, however, find that the spin density within a given Nb$_3$ cluster is largely collinear for all values of $\mathbf{q}$, which implies that each Nb$_3$ cluster may be regarded as a spin-1/2 site in models based on localized spins.
We thus fit the spin-spiral dispersions to a third nearest-neighbor classical Heisenberg model,
\begin{align}\label{eq:third_nearest_neighbor_Heisenberg}
H = -\frac{1}{2} \sum_{i,j} J_{ij} \mathbf{S}_{i}\cdot\mathbf{S}_j,
\end{align}
where $\mathbf{S}_i$ represents spin-1/2 vectors corresponding to a Nb$_3$ cluster indexed by $i$.
The fits are provided as dashed lines in (a) and are in excellent agreement with the calculated points.
The model (sites represented by small triangles) is depicted in (b) with first nearest-neighbor coupling $J_1$ in blue, second nearest-neighbor $J_2$ in gray, and third nearest-neighbor $J_3$ in purple. Here $J_k$ represent all equivalent $J_{ij}$ connecting the $k$'th nearest neighbors.

The computed exchange parameters allow us to estimate the Weiss temperature\cite{Johnston2015} as $\Theta_w = S (S+1) J_0 / 3 \,k_\mathrm{B}$ with $J_0=\sum_i J_{0i}$.
%
We list our best estimates of $J_n$ and $\Theta_w$ in Table \ref{table:tab2} for the three compounds.
We find a negative $J_1$ for $\mathrm{Nb}_3\mathrm{Cl}_8$ but a positive $J_1$ for the Br and I compounds.
Our value of $\Theta_w = -70.09$ K for $\mathrm{Nb}_3\mathrm{Cl}_8$ is negative and is in agreement with the sign of the experimental value for the bulk material ($\Theta_w = -23.0$ K)\citen{Pasco2019}. We do, however, find a somewhat lower value, which could either point to our calculations overestimating the magnitude of the exchange constants or that the magnetic properties of the monolayer deviate from that of the bulk compound. In any case, the negative experimental value implies the presence of antiferromagnetic exchange interactions.

\begin{table}[t!]
\begin{ruledtabular}
\begin{tabular}{c c c c c}
    & $\mathrm{Nb}_3\mathrm{Cl}_8$ & $\mathrm{Nb}_3\mathrm{Br}_8$  &  $\mathrm{Nb}_3\mathrm{I}_8$ \\
   $J_1$ & -3.438 &   2.123     &   2.753    & \\
   $J_2$ & -0.173 &  -0.423      &   -0.913    & \\
   $J_3$ & -0.416 &  -0.539      &  -1.261     & \\
   $\Theta_w$ & -70.09     &  20.21      &    10.07   & \\
   $m^\mathcal{S}$ &  0.199  &  0.197      &  0.192     & \\
   $m^\mathcal{L}$ &  0.008    &    0.016    &   0.040    & \\
\end{tabular}
\end{ruledtabular}
\caption{\label{table:tab2} Magnetic properties of the ground states. Exchange constants $J_n$ are provided in units of meV and the Weiss temperatures are given in units of Kelvin. Spin ($\mathcal{S}$) and orbital ($\mathcal{L}$) moments (in $\mu\mathrm{B}$) are reported for the FM state and is obtained by integration within the PAW sphere of a Nb site. 
}
\end{table}

To verify the spin-spiral calculations, we cast the relaxed structures into $\sqrt{3}\times\sqrt{3}$ super cells containing three formula units. 
%
%
We then perform a non-collinear calculation with spin-orbit coupling (SOC) included self-consistently (SOC were excluded from the spin spiral calculations). 
%
%
For the case of $\mathrm{Nb}_3\mathrm{Cl}_8$, the $\mathbf{Q} = (\frac{1}{3}, \frac{1}{3})$ cycloid [displayed in Fig.~\ref{fig2}~(d)], has an energy which is lower by $4.37$ meV/f.u.~compared to the FM state, agreeing very well with the spin-spiral calculation ($\Delta E$ = 4.32 meV).  
We do not investigate the long-period helical phases of the Br and I systems, as those would require a computationally intractable super cells, but we do find that a FM state has lower energy than any spin orientation along high-symmetry BZ directions by around 2 meV/f.u thus agreeing well with the spin spiral calculations. This also shows that it would be a highly non-trivial task to determine the correct magnetic ground state without using the generalized Bloch theorem and spin spiral calculations.
Curiously, in all cases where SOC is included, we find the imposed FM configuration to be weakly noncollinear - shown in Fig.~\ref{fig2}~(c). The individual Nb spins cant from the out-of-plane direction towards the plane of the monolayer. 
For Cl, Br, and I, the values of the polar angles are $\theta^\mathrm{nc} = 3.9^\circ, 6.1^\circ$ and $10.2^\circ$ respectively.
We estimate the orbital magnetic moment of this configuration using the atom-centered approach outlined in Ref.~[\citen{Ovesen2024}] and reveal that it is weak (nearly quenched).
In the canted FM state, the in-plane components of the orbital moment are larger than the in-plane spin components. For the iodine compound we find $m_x^\mathcal{L}/m_x^\mathcal{S} = m_y^\mathcal{L}/m_y^\mathcal{S} \simeq -1.2$ whereas for the primary component $m_z^\mathcal{L}/m_z^\mathcal{S} \simeq -0.04$. Similar trends are observed for the other compounds implying that the spin and orbital moments are not aligned. 
%
%
We list the magnitudes of $m^\mathcal{S}$ and $m^\mathcal{L}$ in Table.~\ref{table:tab2}.
%
%
It should be stressed that the $m^\mathcal{S}$ of the PAW centered approximation sum to $0.6$ $\mu\mathrm{B}$ in the $[\mathrm{Nb}_3]^{8+}$ cluster but a total cell integration of the spin density finds $M^\mathcal{S} = 1.0$ $\mu\mathrm{B}$ as expected for an insulating $S = 1/2$ system. 
This observation suggests large interstitial contributions in the $\mathrm{Nb}_3$ cluster which is typical of a delocalized molecular orbital.
To this end we note that Grytsiuk \emph{et al} [\citen{Grytsiuk2024}] also predicted a $120^\circ$ cycloid state as the magnetic ground state of monolayer $\mathrm{Nb}_3\mathrm{Cl}_8$ from dynamical mean field theory. In that work it was estimated to reside 7 meV below the FM state, which is in reasonable agreement with the present LDA results.
Finally, we note that the magnetic point group derived from $P3m1$ of course depends on the predicted magnetic order. For a FM (collinear) state, the magnetic space group is preserved as $P3m1$ if the magnetic moments are orthogonal to the mirror plane but becomes modified to $P3m'1$ if the moments are parallel to the plane (for example in the out-of-plane direction). For a helical state with the spiral plane coinciding with the atomic plane the symmetry ($P3m1$ or $P3m'1$) depends in principle on the phase of the spiral (this is also the case for a FM state aligned in the plane). However, due to the three-fold rotational symmetry, any in-plane anisotropy effects must vanish to second order in spin operators (referring to Heisenberg-type of models). Thus for the model ~\eqref{eq:third_nearest_neighbor_Heisenberg} one would need terms to fourth order to represent in-plane anisotropy and neglecting such effects imply that symmetry constraints imposed by {\it either} $P3m1$ or $P3m'1$ are approximately satisfied for any type of in-plane order.

\subsection{\label{subsec:lin_me}Linear magnetoelectric response}
As noted above, the character of the magnetic ground state may strongly influence various derived physical properties. Here we exemplify this by analyzing
%
the 2D linear magnetoelectric tensor ${\bm \alpha}^\mathrm{2D}$ whose components are
\begin{align}\label{eq:lin_ME}
\alpha^\mathrm{2D}_{ij} = \mu_0 \left(\frac{\partial M_i}{\partial \mathcal{E}_j}\right) = \mu_0 \left[ \left(\frac{\partial M^\mathcal{S}_i}{\partial \mathcal{E}_j}\right) + \left(\frac{\partial M^\mathcal{L}_i}{\partial \mathcal{E}_j}\right)\right].
\end{align}
%
%
Here, $\mathbf{M}^{(\mathcal{S}, \mathcal{L})} = A^{-1} \sum_a \mathbf{m}_a^{(\mathcal{S}, \mathcal{L})}$ corresponds to the spin $(\mathcal{S})$ and orbital $(\mathcal{L})$ magnetization per unit area with the moment $\mathbf{m}_a^{(\mathcal{S}, \mathcal{L})}$ at a Nb site (indexed by a) and the sum runs over the Nb atoms in the unit cell.
The quantity $A$ is the area of the monolayer in the unit cell and $\mu_0$ the usual permeability of vacuum.
Using the methods outlined in our recent work [\citen{Mangeri2024}], we compute the ionic contributions to Eq.~(\ref{eq:lin_ME}), under a static and homogeneous electric field $\mathbfcal{E}$. This is obtained by considering the ionic displacements introduced by the field\cite{Wu2005, Iniguez2008, Ye2014},
\begin{align}\label{eq:frozenE}
u_\beta(\mathbfcal{E}) = A^{-1}\sum_{j\kappa}\left(C_{\beta\kappa}\right)^{-1} Z_{j\kappa}^e \mathcal{E}_j
\end{align}
Here, $\mathbf{C}$ is the force constant matrix whose pseudoinverse is computed by the Moore-Penrose technique\cite{Strang80} to trace out the acoustic modes. 
The Born effective charges (BECs) $Z_{j\kappa}^e = A \,\partial P_j / \partial u_\kappa$ are calculated from finite differences (in atomic coordinates) with the polarization obtained from the Berry phase approach.
Greek symbols span the atomic coordinates ($\beta = 1,...,3N_a$) and Latin symbols index the Cartesian reference frame $j = x,y,z$. The magnetoelectric tensor is thus obtained by recording the change in magnetization under a varying electric field [using the displacements given in Eq.~\eqref{eq:frozenE}] and fitting a linear function to the result.
One should note that other contributions to Eq.~(\ref{eq:lin_ME}) are nonzero in general, resulting from a purely electronic (clamped-ion) response\cite{Bousquet2011, Malashevich2012} but we do not investigate them here.
The atomic coordinates of the ground state were displaced using Eq.~(\ref{eq:frozenE}) in all three Cartesian unit directions with a maximum amplitude of $\mathcal{\bm E}$ of 0.05 V/nm.
The BECs were computed with maximum displacements of $5\times 10^{-3}$ \AA.
For the FM states (with out-of-plane moments), the space group is $P3m'1$ (No. 156.51), and the magnetoelectric tensor is constrained to the form:
\begin{align}\label{eq:p3bm1}
{\bm \alpha}^\mathrm{2D} (P3m'1) = \begin{pmatrix}
\alpha_{xx}^\mathrm{2D} & 0 & 0\\
  &  & \\
0 & \alpha_{xx}^\mathrm{2D} & 0\\
  &  & \\
0 & 0 & \alpha_{zz}^\mathrm{2D} \\
\end{pmatrix}.
\end{align}
%
%
%
%
The components are obtained from the total magnetization ($\mathcal{S}+\mathcal{L}$) and yields $\alpha_{xx}^\mathrm{2D} = 2.20, \alpha_{zz}^\mathrm{2D} = -0.16$ for Nb$_3$Cl$_8$ and $\alpha_{xx}^\mathrm{2D} = -0.17, \alpha_{zz}^\mathrm{2D} = 0.01$ for Nb$_3$Br$_8$. All values are in units of $\,10^{-3} \,\mathrm{attoseconds}$.
%
%
We note that the convergence of these calculations can be somewhat tricky and the accuracy is on the order of $10^{-5}$ attoseconds (requiring that the maximum absolute change in the electronic density is less than $10^{-10}$ electrons/valence $e^-$ between self-consistent field cycles).
A decomposition of orbital and spin components demonstrates that the lattice-mediated contribution to the linear ME effect is primarily driven by the spin (i.e. the ratio $\alpha_{xx}^{\mathrm{2D},\mathcal{L}}/\alpha_{xx}^{\mathrm{2D},\mathcal{S}} = -0.08$ in the case of Nb$_3$Cl$_8$).
We have checked that ${\bm \alpha}^\mathrm{2D} \to -{\bm \alpha}^\mathrm{2D}$ upon reversal of either polarization state magnetization as expected from the transformation properties of linear magnetoelectric coupling.
Interestingly, $\mathrm{Nb}_3\mathrm{Br}_8$ exhibits a order of magnitude \emph{lower} ME response (and opposite sign) than that of $\mathrm{Nb}_3\mathrm{Cl}_8$. This is surprising since the magnetoelectric effect is driven by SOC and Br is expected to induce significant more SOC compared to Cl. But evidently, the ME response is governed by more elusive effects in these compounds.
We now turn to the issue of the cycloidal ground state in $\mathrm{Nb}_3\mathrm{Cl}_8$. If the phase of the spiral is such that a single spin is orthogonal to the mirror plane we have $P3m1$ and
\begin{align}\label{eq:p3m1}
{\bm \alpha}^\mathrm{2D} (P3m1) = \begin{pmatrix}
0 & \alpha_{xy}^\mathrm{2D}  & 0\\
  &  & \\
-\alpha_{xy}^\mathrm{2D} & 0 & 0\\
  &  & \\
0 & 0 & 0 \\
\end{pmatrix}.
\end{align}
However, if a single spin is parallel to the mirror plane the symmetry becomes $P3m'1$. Due to the in-plane isotropy it is thus likely that any in-plane spiral will have a magnetoelectric tensor confined by both $P3m1$ and $P3m'1$, which implies that all components must vanish. 
The computations using Eq.~(\ref{eq:lin_ME}) yield $|\alpha_{ij}^\mathrm{2D}| \lesssim $ $\,10^{-5}\,$ in units of attoseconds which are significantly lower than for the FM state and are at or below the precision of our calculations. 
This confirms that the linear ME effect vanishes for a spin cycloid.
We should clarify that the units (attoseconds) of ${\bm \alpha}^\mathrm{2D}$ are different than for bulk crystals, which are typically provided in S.I. units of $\mathrm{ps}\cdot\mathrm{m}^{-1}$. We refer the reader to Ref.~[\citen{Rivera2009}] for a discussion of the units of and simply note that the values may be converted into standard 3D magnetoelectric coupling coefficients by dividing by a typical bulk layer distance of $6-8$ \AA. For the FM state of $\mathrm{Nb}_3\mathrm{Cl}_8$ this conversion yields a transverse component of a 3-4 $\mathrm{ps}\cdot\mathrm{m}^{-1}$, which is comparable to the bulk value of Cr$_2$O$_3$\cite{Mangeri2024}.
\subsection{\label{sec:bands} Electronic bands of $\mathrm{Nb}_3\mathrm{Cl}_8$}

\vspace*{-10pt}

\begin{figure}[t!]\centering
\includegraphics[height=12.25cm]{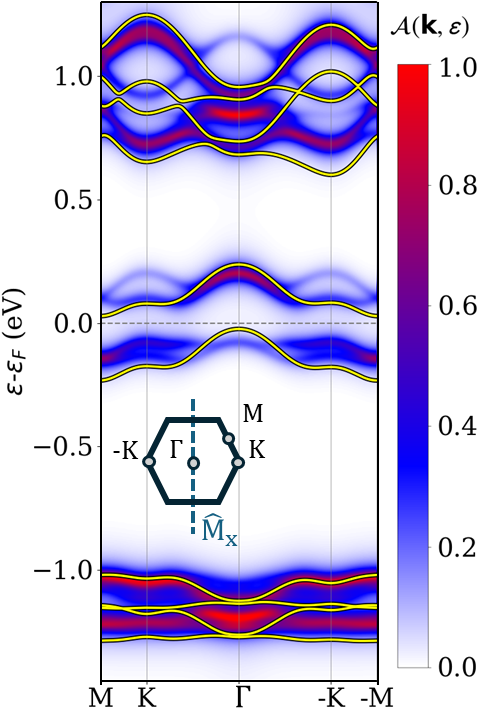} 
\caption{\label{fig3} Electronic band structure of $\mathrm{Nb}_3\mathrm{Cl}_8$ for the out-of-plane FM spin state (solid yellow lines) compared with the spectral function $\mathcal{A}(\mathbf{k},\varepsilon)$ unfolded from the helical phase represented in a super cell. Spin-orbit coupling is included self-consistently in both calculations.}
\end{figure}

Monolayers of $\mathrm{Nb}_3\mathrm{X}_8$ have received interest as valleytronic materials \cite{Peng2020, Feng2023}, displaying non-reciprocal electronic bands along the $\Gamma\to\pm $K coordinates in the BZ.
Indeed, one finds non-reciprocity in the out-of-plane FM state (primitive unit cell) owing to the broken time-reversal and inversion symmetry (solid yellow lines in Fig.~\ref{fig3}). 
The non-reciprocity may be quantified by the eigenvalue differences $\varepsilon_{n,\mathrm{K}}-\varepsilon_{n,\mathrm{-K}}$ where $n$ is a band index such as the lowest conduction band (c) or highest valence band (v).
We find for example $\varepsilon_{(c+1),\mathrm{-K}}-\varepsilon_{(c+1),\mathrm{K}} = 43$ meV and $\varepsilon_{(v-1),\mathrm{-K}}-\varepsilon_{(v-1),\mathrm{K}} = 18$ meV.
The reversal of either the spin or the out-of-plane polarization interchanges the eigenvalues as $\mathrm{K} \to \mathrm{-K}$, thus multiple valley configurations are accessible as demonstrated for $\mathrm{Nb}_3\mathrm{I}_8$ in Ref.~[\citen{Peng2020}].
As discussed in Sec.~\ref{subsec:mag}, both spin-spiral and self-consistent DFT calculations, yield the $120^\circ$ in-plane cycloid as the ground state for $\mathrm{Nb}_3\mathrm{Cl}_8$.
In order to compare the band structure with that of the FM states, we unfold the commensurate spiral super cell eigenvalues and wavefunctions into the BZ of the primitive cell and compute the spectral function defined by
\begin{align}\label{eq:spec}
\mathcal{A}(\mathbf{k},\varepsilon) = \frac{\eta}{\pi} \sum\limits_m \frac{P_{m,\mathbf{K}}(\mathbf{k})}{(\varepsilon-\varepsilon_{m,\mathbf{K}})^2 + \eta^2} 
\end{align}
where the weights,
\begin{align}\label{eq:weights}
P_{m,\mathbf{K}}(\mathbf{k}) = \sum\limits_n \left|\langle m \mathbf{K}| n \mathbf{k} \rangle\right|^2,
\end{align}
describe how much of the super cell band character $(m\mathbf{K})$ is preserved in the primitive cell representation $(n\mathbf{k})$. 
The states $m\mathbf{K}$ in Eq.~(\ref{eq:spec}) and (\ref{eq:weights}) are uniquely mapped to $n\mathbf{k}$ using the method of Popescu and Zunger\cite{Popescu2012}.
The parameter $\eta$ defines a broadening included for visualization and is set to $27$ meV.
%
%
%

%
%
The spectral function is shown in Fig.~\ref{fig3} and it is clear that the FM non-reciprocity is absent from the cycloid state
%
%
We verify this by evaluating the difference of $\mathcal{A}(\mathrm{K},\varepsilon)-\mathcal{A}(\mathrm{-K},\varepsilon)$ for all $\varepsilon$ and find that it is practically zero.
%
%
%
The degeneracy between states at $\Gamma\to\pm K$ originates from the $m$ symmetry in the $P3m1$ magnetic space group. There are three such mirror planes and in the inset of Fig.~\ref{fig3} we show one of these (M$_\mathrm{x}$) enforcing degeneracy between $k_x$ and $-k_x$. As mentioned above the exact presence of this symmetry depends in principle on the phase of the cycloid state, but due the in-plane isotropy the band structure becomes independent of the phase. In the FM state the $m$ symmetry becomes $m'$ which does not enforce degeneracy between $k_x$ and $-k_x$ and non-reciprocity becomes allowed.
%

%
%
%
In order to verify this we performed calculations of the FM state where the spins were enforced to point in varying directions within the plane. We find that the reciprocity is exactly restored at the point where the spins are aligned with the normal of M$_\mathrm{x}$ such that the $m$ rather than $m'$, magnetic symmetry is realized.
%
%
%

%
\section{\label{sec:switching} Electric polarization}

\vspace*{-10pt}

\subsection{\label{subsec:polar} Polar order}

\vspace*{-10pt}

\begin{figure}[t!]\centering
\includegraphics[height=7.1cm]{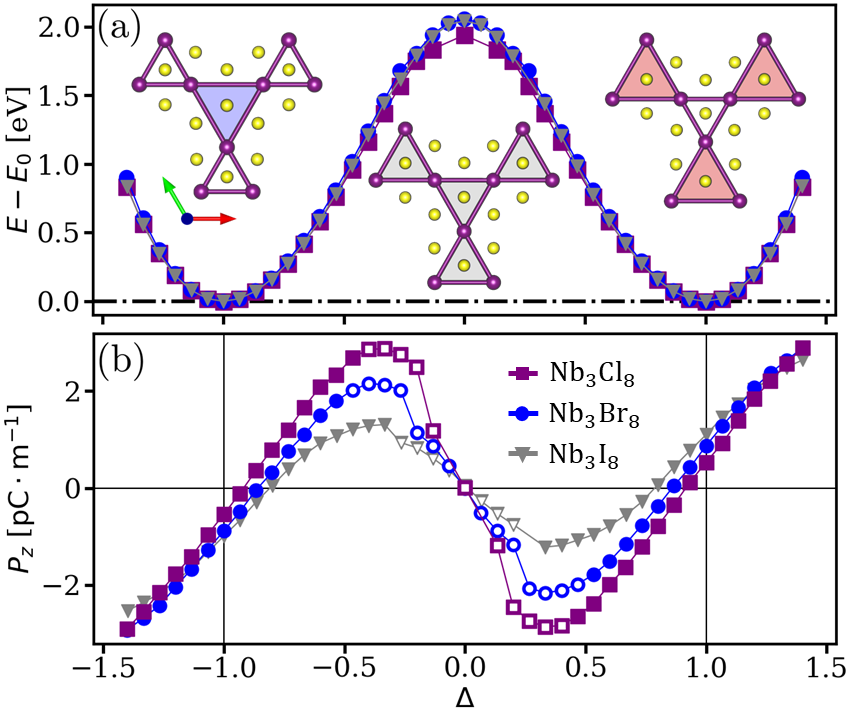} 
\caption{\label{fig4} (a): Energy dependence of the breathing kagom\'{e} coordinate $\Delta$ relative to the ground state. The insets show the structure for $\Delta = -1, 0,$ and $1$ from left to right. Halogens are in yellow and niobium atoms in purple. (b): $P_z$ as a function of the breathing coordinate. The hollow markers indicate metallic phases.}
\end{figure}

Next, we investigate the possibility of switching the state of polarization by application of a homogeneous electric field.
We first use the nudged elastic band (NEB) method with fixed lattice constants to identify the transition state between up and down orientations of the out-of-plane dipole moment.
The transition state is found to be a pure kagom\'{e} lattice with the Nb-Nb bond length equal to half of the lattice constant $(a/2)$.
%
%
%
This atomic configuration has the space group $P\bar{3}m1$ (No. 164) which is inversion symmetric.
%
%
From the NEB minimum energy path, we find that the relevant order parameter of the structural switch is the breathing kagom\'{e} coordinate.
This is quantified as the unitless parameter
\begin{align}\label{eq:breathing}
\Delta = \frac{\delta - \gamma}{\delta_0 - \gamma_0}
\end{align}
with $\delta, \gamma$ being the bond lengths of the small and large triangles and $\delta_0, \gamma_0$ the ground state values as shown in Fig.~\ref{fig1}~(b).
Therefore, when $\Delta$ transits from -1 to  1, it corresponds to flipping the polarization from negative to positive orientations.
We then freeze in the Nb positions (as well as the lattice vectors) and increase $\Delta$ in small increments from -1.4 to 1.4.
All other degrees of freedom (internal coordinates of the halogens) are relaxed with PBE using the dipole layer correction.
This interpolation provides the characteristic double well potential as shown in Fig.~\ref{fig4}~(a).
The data for the Cl, Br, and I compounds are presented in purple squares, blue circles, and gray triangles respectively.
The inset shows the kagom\'{e} structure at $\Delta  = -1$, 0, and 1 from left to right.
The data demonstrates that the energy barrier, around 2.0 eV/f.u. is relatively invariant to the halogen substitution and is in agreement with a previous calculations of the barrier\cite{Feng2023}.
%

For each calculation, we evaluate the out-of-plane 2D polarization $P_z$ by integrating the all-electron charge density $\rho(\mathbf{r}; \Delta)$ (including ionic contributions) at a given $\Delta$,
\begin{align}\label{eq:polarization_z}
P_z(\Delta) = A^{-1} \int z\rho(\mathbf{r}; \Delta)d\mathbf{r}.
\end{align}
This is shown in Fig.~\ref{fig4}~(b) and reveals a peculiar dependence upon the switching parameter $\Delta$.
We have verified that roughly identical results are obtained from the Berry phase approach\cite{King-Smith1993, Resta1994}, but for the present case of out-of-plane polarization in a 2D material, it is easier to apply Eq.~\eqref{eq:polarization_z} since it only requires knowledge of the electronic charge density (rather than the wavefunctions).
At $\Delta = -1$, $P_z$ is negative in all three compounds.
As $\Delta$ evolves towards zero (the nonpolar phase), $P_z$ changes sign at around $\Delta \simeq -0.9$ and continues to grow larger and positive until around $\Delta \simeq -0.5$ after which it decreases again through zero at $\Delta = 0$.
This dependence is antisymmetrically repeated as $\Delta$ evolves from zero to one. In particular, it should be noted that for the case of Nb$_3$Cl$_8$ the magnitude of polarization reaches a maximum in the vicinity of $\Delta=0.4$ where it is more than 5 times larger (and have opposite sign) compared to the ground state spontaneous polarization at $\Delta=1$.
%
%

%
%

%
As $\Delta \to 0$, 
the lattice becomes pure kagom\'{e} and a lack of distinction arises between long and short Nb-Nb bond lengths. 
Therefore, the localization picture of the $\mathrm{Nb}_3$ trimerized spin-1/2 breaks down and a metallic state appears as a consequence of the democratization between equivalent choices of Nb$_3$ trimers.
The metal-insulator transition happens in the vicinity of $\Delta\sim\pm0.5$ and metallic states are marked by open symbols in Fig.~\ref{fig4}~(b).
Such emerging metal-metal bonds have been proposed in other kagom\'{e} compounds\cite{Kelly2019} or in molecular $\mathrm{Fe}^{3+}$ complexes\cite{Sanchez2016} and would usually preclude the possibility of switching the polar state.
%
%
However, in the case of atomically thin materials, the thickness is typically not large enough to completely screen an external electric field and ferroelectric switching may be allowed for out-of-plane polarizations - even if the transition path passes through a metallic state.
This has been experimentally demonstrated in metallic $\mathrm{WTe}_2$ in a seminal paper by Fei \emph{et al}\cite{Fei2018}.
%
%
%
%

While the metallic transition state does not exclude the possibility of ferroelectric switching in the present case, the polarization profile in Fig.~\ref{fig4}~(b) does.
This is easily seen from the electrical enthalpy\cite{Stengel2009,Kruse2023},
\begin{align}\label{eq:enthalpy}
\mathcal{F}(\Delta,\mathbfcal{E}) = E(\Delta) - A \,\mathbf{P}(\Delta)\cdot\mathbfcal{E}
\end{align}
where $E$ is the electronic energy from DFT calculation [shown in Fig.~\ref{fig4}~(a)] and $\mathbfcal{E}$ is the electric field.
The effect of an applied electric field is thus to add a term linear in $\mathbf{P}$ to the double well potential. If the polarization has a monotonous dependence on $\Delta$ it is easy to see that one may find an electric field that makes the electric enthalpy decrease monotonously from one polarization state to the other. However, the zero points of $P_z(\Delta)$ define points in $E(\Delta)$ that are invariant to an applied electric field and these fixed points imply that Eq.~\eqref{eq:enthalpy} cannot become a  monotonous function for any $\mathbfcal{E}$ and the compounds thus cannot be switched.
%

%
The unusual polarization profile originates from the fact the primary order parameter $\Delta$ only involves the position of Nb atoms in the plane while the polarization is strictly out-of-plane. The polarization thus arises from the relative movement of halide atoms which are pushed away or attracted to the center of the layer as the symmetry between equivalent triangles is broken.
%
%
Since the polarization seems to be a associated with a secondary mode (the displacement of halide atoms) it is tempting to assign the behavior derived from Fig.~\ref{fig4} to an improper ferroelectric, which is exactly characterized by a non-polar primary order parameter\cite{Levanyuk1974, Fennie2005, Stengel2012, Garrity2014}.
%
%
Additionally, a cubic dependence of the dipole moment on the switching coordinates has been observed in the improper ferroelectric yttrium manganite\cite{Fennie2005, Garrity2014}.
%
%
%
To investigate the possibility of improper ferroelectric further, we computed the spin-paired $\Gamma$-point phonon spectra of the nonpolar (metallic) phase at $\Delta = 0$,which reveals three unstable phonon modes.
The lowest energy eigenmode is comprised of in-plane relative motions of Nb atoms in the direction of the $\Delta$ coordinate but also contains small out-of-plane action of the halogens.
We analyzed the symmetry of the instabilities using \textsc{spglib}\cite{spglibv1} and conclude that all of them are polar - thus any distortion amplitude or a linear combination of them will break the inversion symmetry.
Therefore, the present compounds cannot be characterized as traditional improper ferroelectrics.
%
%
%

\begin{figure*}[htp!]\centering
\includegraphics[height=7.225cm]{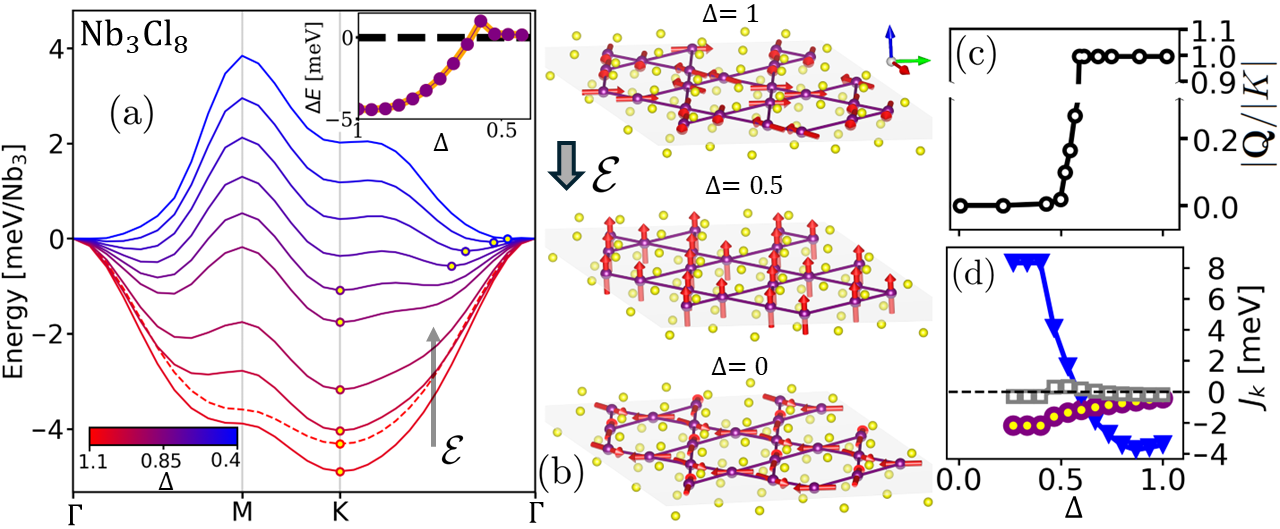} 
\caption{\label{fig5} (a) Spin spiral energies at various values of $\Delta$ reveals a first order phase transition for $\mathbf{Q}=(\mathrm{K})$ to a long wavelength spiral around $\Delta=0.5$. (b) The magnetic states (in a super cell representation) for $\Delta = 1$ (top), $\Delta = 0.5$ (middle) and $\Delta = 0$ (bottom). (c) The ground state ordering vector as a function of $\Delta$ - note the broken $y$-axis signifying the first order phase transition at $\Delta\sim0.5$. (d)
The fitted exchange constants $J_1$ (blue triangles), $J_2$ (gray squares) and $J_3$ (purple circles) across the same range are shown in (d).
%
%
%
}
\end{figure*}

\vspace*{-10pt}

\subsection{\label{subsec:firstord} Electric-field driven magnetic phase transition in $\mathrm{Nb}_3\mathrm{Cl}_8$}

\vspace*{-10pt}

Despite the fact that the materials cannot be coherently switched from positive to negative out-of-plane $P_z$ orientations, they still exhibit strong coupling between magnetic and electric degrees of freedom. 
In order to demonstrate this, we have computed the spin-spiral energy dispersions for a range of $\Delta$ values between $1.1$ and $0.5$ for the case of $\mathrm{Nb}_3\mathrm{Cl}_8$.
These are shown in Fig.~\ref{fig5}~(a) and it is clear that the energy minima stays at the K point in the range of $\Delta=1-0.5$ after which the ordering vector exhibits a discontinuity and jumps to finite (but small) value of $\mathbf{Q}$ corresponding to a long range incommensurate spin spiral. The ground state ordering vector then decreases rapidly (but continuously) and finally settles at $\mathbf{Q}=(0,0)$ corresponding to FM order at $|\Delta|<0.4$.
Thus, the out-of-plane electric field $\mathbfcal{E}$ which couples to $\Delta$ can drive a magnetic phase transition from the helical phase to the out-of-plane FM texture as shown graphically in Fig.~\ref{fig5}~(b) for $\Delta = 1$ and $\Delta = 0.5$.
This is quantified in panel (c) which tracks the location of the spin spiral energy minimum as a function of $\Delta$.
For each each spin-spiral dispersion shown in (a), we fit to the Heisenberg model of Eq.~(\ref{eq:third_nearest_neighbor_Heisenberg}).
This allows one to extract $J_1$, $J_2$, and $J_3$ as functions of the breathing coordinate $\Delta$.
These are shown in (d), which demonstrates that the nearest-neighbor exchange ($J_1$, solid purple line) is negative until around $\Delta \simeq 0.5$ after which it becomes positive and its strength increases monotonically. This drives the transition to the FM state as $\Delta$ decreases.
%
%
As a check that this picture is preserved when SOC is included, we have performed self-consistent SOC calculations along the switching path by considering both FM and $\mathbf{Q}$ order in the $\sqrt{3}\times\sqrt{3}$ super cell. 
The energy difference $\Delta E(\Delta) = E_\mathbf{Q}(\Delta)-E_\mathrm{FM}(\Delta)$ is shown in the inset of (a) and agrees with the spiral calculations (for example exhibiting a sign change in the vicinity of $\Delta \simeq 0.5$). 
%
%
%

Although the $\Delta=0$ state seems to be inaccessible by an external field, it is worth commenting on the magnetic ground state favored in the perfect kagome lattice. Indeed, we find that the trimerization of Nb spins \emph{breaks} at around $\Delta \simeq \pm 0.5$ concomitant with the metal-insulator transition and a new structure emerges with the spins on each Nb site oriented $120^\circ$ to one another in the plane of the monolayer.
This state is shown in the bottom panel of Fig.~\ref{fig5}~(b) for $\Delta = 0$.

\vspace*{-10pt}

\section{\label{sec:concl}Discussion and Outlook}

\vspace*{-10pt}

We investigated the magnetoelectric properties of niobium halide monolayers $\mathrm{Nb}_3\mathrm{(Cl, Br, I)_8}$ using DFT calculations.
In all cases except close to the transition state, we find the spins in the $\mathrm{Nb}_3$ cluster are collinear in accordance with the picture of trimerization and a magnetic model where each cluster is regarded as a localized spin-1/2 site.
For $\mathrm{Nb}_3\mathrm{Cl}_8$, the magnetic interactions are dominated by a nearest neighbor antiferromagnetic exchange interaction between trimers, which leads to a ground state with a $120^\circ$ spin cycloid. Such a state could be directly verifiable using the magneto-optical Kerr effect\cite{Hu2024}.
For $\mathrm{Nb}_3\mathrm{(Br,I)}_8$, our spin-spiral calculations reveal that the ground states are long wavelength incommensurate planar spirals, which have FM nearest neighbor trimer interactions and the spiral states are thus driven by second and third nearest neighbor antiferromagnetic interactions.
%
%
%
%
The symmetry of the spiral ground states imply that the proposed AVHE (based on FM ground states) in these materials is in fact forbidden. This was verified explicitly using unfolded band structure calculations for the case of $\mathrm{Nb}_3\mathrm{Cl}_8$, but also holds true for the other compounds.
%

%
While the monolayers are definitely polar (pyroelectric), we have raised serious doubt on whether the previously predicted ferroelectric assignment is correct. By evaluating the polarization along the switching path it becomes clear that the multiple change in direction of polarization along the path prohibits direct switching by an external field. At least for the case of coherent monodomain switching. Nevertheless, while it is not possible to switch the polarization, one may drive the polar state reversibly by an applied electric field and such change is accompanied by a magnetic phase transition. Thus, although the compounds are non-switchable, the magnetic ground state may be changed by an electric field, which implies that one may control whether the AVHE and ME effects are present or absent. We only showed this explicitly for the case of $\mathrm{Nb}_3\mathrm{Cl}_8$, but by comparing Fig.~\ref{fig2} and \ref{fig5} we believe it to be highly plausible that an electric field may drive $\mathrm{Nb}_3\mathrm{(Br, I)_8}$ into FM states as well. These compounds will, however, not exhibit the first order phase transition from the $120^\circ$ spin cycloid to a long wavelength spiral state characteristic of $\mathrm{Nb}_3\mathrm{Cl}_8$.

At first sight the present work might appear somewhat negative in its conclusions since we have argued that the 2D Nb halides are not switchable and that the proposed AVHE becomes forbidden if the magnetic ground states are correctly accounted for. However, we do predict these materials to be strongly magnetoelectric. Although the linear ME vanishes in the spiral ground states, one may induce a magnetic {\it phase transition } by an external electric field and this in turn reintroduces the AVHE and ME in the field driven state. Moreover, the conclusion regarding non-switchability is solely based on the concept of coherent monodomain switching. In reality, the mechanism responsible for switching in most ferroelectrics involve domain nucleation and/or domain wall propagation \cite{Petralanda2022}.

Finally we have neglected to discuss the general subtleties associated with magnetic order in 2D materials\cite{Olsen2024}. In particular, the Mermin-Wagner theorem states that magnetic order cannot persist in 2D in the presence of continuous symmetries \cite{Mermin1966}. The in-plane spiral states considered here appears to have exactly such symmetries, since all spins may be rotated freely in the isotropic atomic plane. On the other hand, it has been demonstrated that monolayer NiI$_2$ (also exhibiting a spiral ground state and same crystallographic space group) remains ordered at finite temperatures\cite{Song2022} so it is definitely possible for 2D spiral states to order. Whether or not the exact mechanism is related to higher order single-ion anisotropy terms or other mechanisms will be left for future work. In any case we hope the present work will inspire experimental as well as theoretical scrutiny of the exciting Nb halides in the near future.

\vspace*{-12pt}

\begin{acknowledgments}

\vspace*{-10pt}

All authors acknowledge funding from the Villum foundation Grant No.~00029378. 
Computations were performed on Nilfheim, a high performance computing cluster at the Technical University of Denmark.

\end{acknowledgments}

\bibliography{apssamp}

\end{document}